\documentclass[aps,prd,twocolumn,10pt,amsfonts,amsmath,amssymb]{revtex4-1}

\usepackage{hyperref}
\usepackage{graphicx}
\usepackage{enumerate}

\hypersetup{
  colorlinks   = true, 
  urlcolor     = blue, 
  linkcolor    = blue, 
  citecolor   = blue 
}

\DeclareMathSizes{10}{9}{6}{5}

\newcommand{\be}{\begin{equation}}
\newcommand{\ee}{\end{equation}}
\newcommand{\bea}{\begin{eqnarray}}
\newcommand{\eea}{\end{eqnarray}}
\newcommand{\benn}{\begin{eqnarray*}}
\newcommand{\eenn}{\end{eqnarray*}}
\newcommand{\p}{\phantom}

\newcommand{\dif}{{\rm d}}

\def\bse{\begin{subequations}}%
\def\ese{\end{subequations}}%

\def\apss{Astrophysics and Space Science}
\def\jcap{Journal of Cosmology and Astroparticle Physics}
\def\prd{Physical Review D}

\def\physrep{Physics Reports}

\begin{document}

\title{Relativistic stars in Starobinsky gravity with the matched asymptotic expansions method}

\author{Sava{\c s} Arapo{\u g}lu}
\email{arapoglu@itu.edu.tr}
\affiliation{Istanbul Technical University, Faculty of Science and Letters,  Physics Engineering Department, 34469, Maslak, Istanbul, Turkey}

\author{Sercan {\c C}{\i}k{\i}nto{\u g}lu}
\email{cikintoglus@itu.edu.tr}
\affiliation{Istanbul Technical University, Faculty of Science and Letters,  Physics Engineering Department, 34469, Maslak, Istanbul, Turkey}

\author{K.~Yavuz Ek{\c s}i}
\email{eksi@itu.edu.tr}
\affiliation{Istanbul Technical University, Faculty of Science and Letters,  Physics Engineering Department, 34469, Maslak, Istanbul, Turkey}


\begin{abstract}
We study the structure of relativistic stars in $\mathcal{R}+\alpha \mathcal{R}^{2}$ theory using the method of matched asymptotic expansion to handle the higher order derivatives in field equations arising from the higher order curvature term. We find solutions, parametrized by $\alpha$, for uniform density stars. 
We obtain the mass-radius relations and study the dependence of maximum mass on $\alpha$. We find that $M_{\max}$ is almost linearly proportional to $\alpha$. For each $\alpha$ the maximum mass configuration has the biggest compactness parameter ($\eta = GM/Rc^2$), and we argue that the general relativistic stellar configuration corresponding to $\alpha=0$ is the least compact among these.
\end{abstract}

\maketitle

\section{Introduction}

Modifying Einstein's general relativity (GR) is an old attempt dating back to  Kaluza-Klein's original idea of combining gravity with electromagnetism by introducing a fifth dimension.  
Later the work of Brans-Dicke led to a theory in which a scalar field is coupled to the curvature, and the theory was for many years the unique alternative to GR (see \cite{goenner04} for a review).  Then superstring theory came onto the stage, capturing the idea of Kaluza-Klein and Brans-Dicke in a natural way, appearing to be the most promising theory for unification of gravity and other interactions, and/or quantizing gravity.  In the effective action of such superstring theories, there are ghost-free series of higher curvature corrections to the leading Einstein-Hilbert term representing GR.

The current interest in modifying GR has revived with one of the most important discoveries in physics, namely the current accelerated expansion of the Universe.  A family of modification of GR includes higher curvature theories of the form $\mathcal{R}+\mathcal{R}^n$, where $\mathcal{R}$ is the Einstein-Hilbert term and $\mathcal{R}^n$ corresponds to the $n$th power of curvature scalar, Ricci tensor, Riemann tensor, and Weyl tensors \cite{stel77,stel78,boul83,horo85,deser03,deser07,maldacena11} motivated by candidate fundamental theories.  Another approach for modifying GR is the so-called $f(\mathcal{R})$ theories of gravity \cite{Sotiriou+10,Felice+10,cap11,Odin+11} where one replaces the usual Einstein-Hilbert term in the action with a function of scalar curvature $\mathcal{R}$.   There are many models with different functions of curvature terms, but a 
viable model, while providing accelerated expansion of the Universe \cite{ame+07,avi+13,guo13,Odin+08} (see \cite{cli+12} for a review), should also be consistent
with the solar system and cosmological observations in large scale \cite{cli08}.

A model which can be considered in the intersection of aforementioned modifications of GR is the $\mathcal{R}+\alpha \mathcal{R}^2$ gravity which is physical and does not contain ghostlike modes, unlike the situation encountered when other quadratic curvature terms or some complicated functions of curvature scalar are included in the action.  This $\mathcal{R}+\alpha \mathcal{R}^2$ theory, known as the Starobinsky model \cite{Starobinsky80}, propagates an additional massive spin-0 state in addition to the usual massless graviton, and it was one of the first consistent inflationary models.   It is also of interest to study the implications of this model in
the strong gravity regime by examining the effect of the higher
curvature term on,  for example, the existence and structure of relativistic stars.

Since the field equations of the model are fourth order, unlike the second order field equations of GR, the first attempts to probe the existence of relativistic stars in the Starobinsky model of gravity followed the approach of mapping the model to a scalar-tensor model \citep{Kobayashi+09}.  This approach seems to simplify the analysis by reducing the order of the equations, but may lead to dubious conclusions \cite{Salgado+11}.  Indeed, the first conclusions, reached by applying this method, about the nonexistence of relativistic stars in $f(\mathcal{R})$ theories are corrected by more careful analysis \citep{Bab+09,Bab+10}.

It is, thus, favorable to consider the theory in the originally suggested form without mapping to any equivalent theory; but, to eliminate the difficulties arising from the higher order field equations,  a perturbative approach may need to be invoked.
Such a technique for reducing the order of field equations, known as perturbative constraints or order reduction  \cite{JaenX+86,Eliezer+89}, 
is applied in the strong gravity regime in Ref. \cite{Psaltis+10} to show the existence of relativistic stars in the
Starobinsky model.  
The same method is employed  in Ref. \cite{Eksi+11} for a representative sample of realistic equation of states (EoSs) to constrain the
value of $\alpha$ by using neutron star mass-radius measurements \cite{oze+10}.
A similar work with different results is done in Ref.~\cite{Yaz+14a} by mapping to scalar 
tensor theories with a nonperturbative approach and self-consistent method.
Moreover, the rotating neutron stars are studied with the same method as in Refs.~\cite{Yaz+14b,Yaz+15a}.
 Furthermore, the mass-radius relations are obtained for various $f(R)$ models with realistic EoSs in \cite{Odin+13,Odin+16}, but see \cite{Yaz+15b}.
Besides various $f(R)$ models and containing the Gauss-Bonnet invariant, $f\left(\mathcal{G}\right)$, models are examined by considering hyperons and magnetic field effects in the context of neutron stars in \cite{ast+14,Odin+15,Odi+15}.

Singular perturbation problems where the perturbative term has the highest order of derivative are well known in the study of fluids. A well known example is the viscous term of the Navier-Stokes equation of hydrodynamics which is of second order and increases the spatial order of the Euler equation. The cases where viscosity is small cannot be handled with ordinary perturbative methods in which one would simply ignore the viscous term at the zeroth order as such an equation with reduced order cannot be made to satisfy all boundary conditions. Even if the viscous term could be negligibly small in the bulk of the flow, it would still be the dominant term near the boundary where the flow matches to the given boundary condition in a narrow domain called the boundary layer.  The method of matched asymptotic expansion (MAE)  \citep{bender+78,dyke64} is suitable for handling such singular perturbation problems where the perturbative term has the highest order of derivative. This method in the cosmological setting is employed for a specific class of $f(\mathcal{R})$ theories in Ref. \cite{Evans+08} to handle the higher order derivative terms arising from the terms in the Lagrangian  that are not linear in $\mathcal{R}$. Other than $f\left(\mathcal{R}\right)$ theories, the method is also used in \cite{Shaw+06,Shaw+11,Shaw2+11} for cosmological context.

In this work  we employ the MAE method  for analyzing the structure of neutron
stars in the Starobinsky model of gravity. The method was previously applied for relativistic stars in Ref. \cite{Ganguly+14}
where the authors considered only the trace equation with the MAE method and they solved the hydrostatic equilibrium equations by numerical methods for various equations of states.  
But this necessitates a fine-tuning to match with the Schwarzschild solution at the surface.  We thus apply the method not only to the trace equation but also to the hydrostatic equilibrium equations simultaneously.

The plan of the paper is as follows: In Sec.\ \ref{sec:setup}, the field equations and the hydrostatic equilibrium equations are obtained.  In Sec.\ \ref{sec:problem}, the MAE method is applied to hydrostatic equilibrium equations, and in Sec.\ \ref{sec:unistars}, we obtain the solutions for uniform density stars. In Sec.\ \ref{sec:mr} we obtain the mass-radius relations (depending on $\alpha$) in this gravity model and examine the maximum mass depending on $\alpha$. Finally, in Sec.\ \ref{sec:conc} we present our conclusions.

\section{Field Equations and Setup}
\label{sec:setup}
The action of the Starobinsky model is
\begin{equation}
S=\frac{1}{16\pi }\int {\rm d}^{4}x\sqrt{-g}\left(\mathcal{R}+\alpha \mathcal{R}^2\right)+S_{\rm matter},
\end{equation}
where $g$ is the determinant of the metric $g_{\mu\nu}$, $\mathcal{R}$ is the Ricci scalar, and $S_{\rm matter}$ is the matter action. In the metric formalism, the variation of the action with respect to the metric gives the field equations, 
\begin{equation}
\left( 1+2\alpha \mathcal{R}\right) G_{\mu \nu }+\frac{1}{2}\alpha g_{\mu \nu }\mathcal{R}^2
-2\alpha \left( \nabla _{\mu }\nabla _{\nu }-g_{\mu \nu }\square \right)
\mathcal{R}=8\pi T_{\mu\nu }
\end{equation}
\cite{Sotiriou+10,Felice+10}. Contracting with the inverse metric, the trace equation is
\begin{equation}
6\alpha \square \mathcal{R}- \mathcal{R}=8\pi T.
\end{equation}
We assume a  spherically symmetric metric,
\begin{equation}
{\rm d} s^2=-e^{2\Phi}{\rm d}t^2+ e^{2\lambda}{\rm d}r^2 + r^2\left( {\rm d}\theta^2+\sin^2\theta{\rm d}\phi^2 \right),
\end{equation}
where $\lambda=\lambda(r)$ and $\Phi=\Phi(r)$ are the metric functions.
The trace equation, for this form of the spherically symmetric metric, becomes
\begin{align}
6\alpha \exp (-2\lambda )\mathcal{R}^{\prime \prime } =&8\pi T+\left( 1-2\alpha
\mathcal{R}\right) \mathcal{R}+2\alpha \mathcal{R}^2  \notag \\
&+6\alpha \mathcal{R}^{\prime }\exp (-2\lambda )\left( \lambda ^{\prime }-\frac{2}{r}%
-\Phi ^{\prime }\right),  \label{trace1}
\end{align}
where primes denote derivatives with respect to the radial coordinate $r$.
The ``tt'' and ``rr'' components of the field equations are
\begin{align}
-8\pi \rho =&-r^{-2}+\exp (-2\lambda )\left( 1-2r\lambda ^{\prime }\right)
r^{-2} \notag \\
&+2\alpha \mathcal{R}\left( -r^{-2}+\exp (-2\lambda )\left( 1-2r\lambda ^{\prime
}\right) r^{-2}\right)  \notag \\
&+\frac{1}{2}\alpha \mathcal{R}^2 +2\alpha \exp (-2\lambda )\left( \mathcal{R}^{\prime
}r^{-1}\left( 2-r\lambda ^{\prime }\right) +\mathcal{R}^{\prime \prime }\right)
\end{align}
and
\begin{align}
8\pi P=&-r^{-2}+\exp (-2\lambda )\left( 1+2r\Phi^{\prime }\right)r^{-2} \notag \\
&+2\alpha \mathcal{R}\left( -r^{-2}+\exp (-2\lambda )\left( 1+2r\Phi^{\prime
}\right) r^{-2}\right)  \notag \\
& +\frac{1}{2}\alpha \mathcal{R}^2 +2\alpha\exp (-2\lambda ) \mathcal{R}^{\prime }r^{-1}\left(
2+r\Phi ^{\prime }\right),
\end{align}
respectively.

To cast the equations into a more familiar form of the so-called Tolman-Oppenheimer-Volkov (TOV) equations, we define
\begin{equation}
\exp (-2\lambda )=1-\frac{2m(r)}{r}.  \label{lambda_m}
\end{equation}%
We cannot call $m$ as the mass
within the radial coordinate $r$ without determining whether the outside solution is Schwarzschild.
By plugging this to the tt component of
the field equations, the first TOV equation is obtained
as
\begin{align}
&2\left( 1+2\alpha \mathcal{R}+\alpha \mathcal{R}^{\prime }r\right) \frac{{\rm d}m}{{\rm d}r} \notag \\
&\phantom{a}= 8\pi \rho
r^{2}+\frac{1}{2}\alpha r^{2}\mathcal{R}^2 +\alpha \mathcal{R}^{\prime }r\left( -6\frac{m}{r}+4\right) \notag \\
&\phantom{aa}+2\alpha\left( 1-2\frac{m}{r}\right)r^{2}\mathcal{R}^{\prime \prime }.
\label{TOV1}
\end{align}
By using the $r$ component of the
conservation equation of the energy-momentum tensor, $\nabla _{\mu
}T_{\p{1}1}^\mu=0$, we obtain
\begin{align}
\nabla_{\mu}T_{\p{1}1}^\mu=& \partial_{\mu}T_{\p{1}1}^\mu+\Gamma_{\p{1}\mu
\lambda}^{\mu}T_{\p{1}1}^{\lambda}-\Gamma_{\p{1}\mu 1}^{\lambda }T_{\p{1}\lambda}^{\mu}  
\notag \\
=& P^{\prime}+ P (\lambda^\prime + \frac{2}{r}+\Phi^{\prime} ) 
-  (\lambda^\prime P + \frac{2}{r}P - \Phi^\prime\rho ).
\end{align}%
Then the second TOV equation is obtained as 
\begin{equation}
\frac{{\rm d}P}{{\rm d}r}=-\left( \rho +P\right) \Phi ^{\prime },  \label{TOV2}
\end{equation}%
where the $\Phi ^{\prime }$ is found from the rr component of the field equation 
\begin{align}
&2\Phi ^{\prime }\left( 1+2\alpha \mathcal{R}+\alpha r\mathcal{R}^{\prime }\right) \notag \\
&\phantom{a}= 8\pi
P\left( \frac{r^{2}}{r-2m}\right) 
+\left( 1+2\alpha \mathcal{R}\right) \frac{2m}{r-2m}%
r^{-1}  \notag \\
&\phantom{aa}-\frac{1}{2}\alpha \mathcal{R}^2 \frac{r^{2}}{r-2m}-4\alpha \mathcal{R}^{\prime }.
\end{align}

\section{Singular Perturbation Problem \label{sec:problem}}

We, first, define the dimensionless parameters
\begin{align}
x=\frac{r}{R_{\ast}},\quad \epsilon =\frac{\alpha }{R_{\ast}^{2}},\quad \bar{\mathcal{R}}=R_{\ast}^{2}\mathcal{R}, \notag\\ 
\bar{P}=R_{\ast}^{2}P,\quad \bar{\rho}=R_{\ast}^{2}\rho,\quad \bar{m}=\frac{m}{R_{\ast}},  \label{nondim}
\end{align}
where $R_{\ast}$ is the radial distance from center to the surface of the
star and so $0<x<1$. The first TOV equation, Eq.\ \eqref{TOV1}, in terms of these dimensionless
variables becomes
\begin{align}
&\left( 1+2\epsilon \bar{\mathcal{R}}+\epsilon \bar{\mathcal{R}}^{\prime }x\right) \frac{{\rm d}\bar{m}%
}{{\rm d}x}  \notag \\
&\phantom{a}=\frac{x^{2}}{12}\left( 48\pi \bar{P}+\left( 2+3\epsilon \bar{\mathcal{R}}%
\right) \bar{\mathcal{R}}+32\pi \bar{\rho}\right)  \notag \\
&\phantom{aa}+\frac{1}{\left( 1+2\epsilon \bar{\mathcal{R}}\right) }\frac{\epsilon \bar{\mathcal{R}}^{\prime
}}{6}\left[ -6\bar{m}\left( 1+2\epsilon \bar{\mathcal{R}}\right) +x^{3}\left( \bar{\mathcal{R}}%
+3\epsilon \bar{\mathcal{R}}^{2}+16\pi \bar{\rho}\right) \right]  \notag \\
&\phantom{aa}+\epsilon ^{2}x\left( x-2\bar{m}\right) \frac{2}{\left( 1+2\epsilon \bar{\mathcal{R}}%
\right) }\bar{\mathcal{R}}^{\prime {2}} .  \label{dif_m}
\end{align}
Similarly, the second TOV equation, Eq.\ \eqref{TOV2}, then becomes 
\begin{align}
\frac{{\rm d}\bar{P}}{{\rm d}x}=&-\frac{\bar{\rho}+\bar{P}}{4x\left( x-2\bar{m}\right)
\left( 1+2\epsilon \bar{\mathcal{R}}+\epsilon x\bar{\mathcal{R}}^{\prime }\right) } \,  
\notag \\
&\times\left[ 16\pi x^{3}\bar{P}+4\bar{m}+8\epsilon \bar{m}\bar{\mathcal{R}}-\epsilon x^{3}%
\bar{\mathcal{R}}^{2} \right. \notag \\
&\left.-8\epsilon x \bar{\mathcal{R}} ^{\prime }\left( x-2\bar{m}\right) \right] .
\label{dif_P}
\end{align}
Finally, the trace equation Eq.\ \eqref{trace1} becomes 
\begin{align}
&\epsilon \left( 1+2\epsilon \bar{\mathcal{R}}\right) \bar{\mathcal{R}}^{\prime \prime } \notag \\
&\phantom{a}=\left(
-8\pi \bar{\rho}+24\pi \bar{P}+\bar{\mathcal{R}}\right) \frac{1}{6}\left( \frac{x}{x-2%
\bar{m}}\right) \left( 1+2\epsilon \bar{\mathcal{R}}\right)  \notag \\
&\phantom{aa}+\frac{1}{6}\frac{\epsilon }{x-2\bar{m}}\left[ \left( 1+2\epsilon \bar{\mathcal{R}}%
\right) 12\bar{m}x^{-1}-12\left( 1+2\epsilon \bar{\mathcal{R}}\right) \right] \bar{\mathcal{R}}%
^{\prime }  \notag \\
&\phantom{aa}+\frac{1}{6}\frac{\epsilon }{x-2\bar{m}}\left[ 3\epsilon x^{2} \bar{\mathcal{R}}%
^{2}+x^{2}\bar{\mathcal{R}}+16\pi x^{2}\bar{\rho} \right] \bar{\mathcal{R}}^{\prime } +2\epsilon
^{2}\bar{\mathcal{R}}^{\prime {2}}.  \label{dif_cur}
\end{align}
For satisfying continuity at the center of the star we apply two boundary
conditions, $\bar{m}(0)=0$ and $\bar{\mathcal{R}}^{\prime }(0)=0$. Since at the surface
of the star pressure vanishes, we have $\bar{P}(1)=0$.
In general relativity, the Schwarzschild solution is the unique vacuum solution
around a spherically symmetric and static mass distribution according to the Birkhoff theorem 
\cite{wald84, Birk23}. Yet, there is not a unique vacuum solution in $f(R)$ theories \cite{Cik17}. 
In this paper, we examine the case that the Ricci scalar at the surface of the star is just as 
in General Relativity. So, our last boundary condition is $\bar{\mathcal{R}}\left(1\right)=8\pi\bar{\rho}\left(1\right)$. This value is generally zero for realistic equation of states, yet it is nonzero for uniform density distribution.
At this point, we assume that the spacetime outside the star is described by Schwarzschild's metric. 
This assumption allows us to compare our results easily with those of the General Relativity, and also for simplicity. 
Then, with this choice, $m$ denotes the mass within radial coordinate $r$.

Because $\epsilon $ is multiplying $\mathcal{R}^{\prime \prime }$
in the trace equation, Eq.\ \eqref{dif_cur}, the system of equations poses a
singular perturbation problem. An appropriate method for handling such
singular problems, well known in the fluid dynamics research, is the MAE
which we employ in the following. According to the method, there should be a 
\textit{boundary layer} which according to the authors of Ref.\ \cite%
{Ganguly+14} forms near to the surface of the star. The solution which is
valid in the boundary layer is called the \textit{inner solution}, and for
stretching this region a new parameter will be defined in an appropriate
way. Outside the boundary layer---the rest of the star---the \textit{outer
solution} is valid. In a transition region the two solutions match with each
other. The outer solutions of Eqs.\ \eqref{dif_m}, \eqref{dif_P}, and \eqref{dif_cur} 
satisfy the boundary conditions at $x=0$, and the inner solutions
satisfy the boundary conditions at $x=1$.

\section{Uniform density stars}
\label{sec:unistars}

The bulks of neutron stars have a remarkably constant density though the density near to the surface drops by $15$ orders of magnitude. This crustal contribution does not change the mass and the  radius significantly. In this work, we employ the uniform density approach for simplicity.

\subsection{Outer solutions for uniform density}

The outer solutions are valid where $0<x<1$, they are introduced as
perturbative expansions
\begin{subequations}
\begin{align}
\bar{\mathcal{R}}^{\mathrm{out}}(x)=& 
\bar{\mathcal{R}}_{0}^{\mathrm{out}}(x)+\epsilon\bar{\mathcal{R}}_{1}^{\mathrm{out}}(x)
+O\left( \epsilon^2 \right), \\
\bar{m}^{\mathrm{out}}(x)=& 
\bar{m}_{0}^{\mathrm{out}}(x)+\epsilon\bar{m}_{1}^{\mathrm{out}}(x)+O\left( \epsilon^2 \right), \\
\bar{P}^{\mathrm{out}}(x)=& 
\bar{P}_{0}^{\mathrm{out}}(x)+\epsilon\bar{P}_{1}^{\mathrm{out}}(x)+O\left( \epsilon^2 \right),
\end{align}
\end{subequations}
and they should satisfy boundary conditions at $x=0$.
After plugging these expressions into Eqs.\ \eqref{dif_m},  \eqref{dif_P}, and \eqref{dif_cur}, 
the $O(1)$ terms in these equations are
\begin{subequations}
\begin{align}
\frac{{\rm d}\bar{m}_{0}^{\mathrm{out}}}{{\rm d}x} =&\frac{x^{2}}{6}\left( 24\pi \bar{P}%
_{0}^{\mathrm{out}}+\bar{\mathcal{R}}_{0}^{\mathrm{out}}+16\pi \bar{\rho}\right),  \label{out_m_0_diff} \\
4x ( x-2\bar{m}_{0}^{\mathrm{out}} ) \frac{{\rm d}\bar{P}_{0}^{\mathrm{out%
}}}{{\rm d}x} =&- ( \bar{\rho}+\bar{P}_{0}^{\mathrm{out}} ) ( 16\pi x^{3}\bar{P}_{0}^{\mathrm{out}}+4\bar{m}_{0}^{\mathrm{out}} ),  \label{out_P_0_diff} \\
0 =&x\left( -8\pi \bar{\rho}+24\pi\bar{P}_{0}^{\mathrm{%
out}}+\bar{\mathcal{R}}_{0}^{\mathrm{out}}\right).  \label{out_R_0_diff}
\end{align}
\end{subequations}
Equation \eqref{out_R_0_diff} can be written as 
\begin{equation}
\bar{\mathcal{R}}_{0}^{\mathrm{out}}\left( x\right) =8\pi\bar{\rho}
-24\pi \bar{P}_{0}^{\mathrm{out}}\left( x\right) .
\label{Uni_out_R_0_diff2}
\end{equation}%
If we plug it into Eq.\ \eqref{out_m_0_diff}, we obtain
\begin{equation}
\frac{\dif\bar{m}_{0}^{\mathrm{out}}}{\dif x}=4\pi x^{2}\bar{\rho}.
\end{equation}%
The solution of this equation with the boundary condition, $\bar{m}\left(0\right)=0$,
is
\begin{equation}
\bar{m}_{0}^{\mathrm{out}}\left( x\right) =\frac{4}{3}\pi x^{3}\bar{\rho},  \label{Uni_m0}
\end{equation}%
and the solution of Eq.\ \eqref{out_P_0_diff} (see Sec.\ \ref{app_uni_P0}) is 
\begin{equation}
\bar{P}_{0}^{\mathrm{out}}=\bar{\rho}\left[ 
\frac{2\left(P_{c}+\bar{\rho}\right)}
{3\left( P_{c}+\bar{\rho}\right) 
-\left(3P_{c}+\bar{\rho}\right)
\sqrt{1-\frac{8}{3}\pi \bar{\rho}x^2} 
}-1\right].
\end{equation}
According to Eq.\ \eqref{Uni_out_R_0_diff2}, 
\begin{equation}
\bar{\mathcal{R}}_{0}^{\mathrm{out}}\left( x\right) =
16\pi \bar{\rho}
\left[ 1-
\frac{ \left(3P_{c}+\bar{\rho}\right)\sqrt{1-\frac{8}{3}\pi \bar{\rho}x^2}}
{3\left( P_{c}+\bar{\rho}\right)
-\left(3P_{c}+\bar{\rho}\right)\sqrt{1-\frac{8}{3}\pi \bar{\rho}x^2}}
\right].
\end{equation}

After plugging the outer solutions into Eqs. \eqref{dif_m},  \eqref{dif_P}, and \eqref{dif_cur},
the $O(\epsilon)$ terms in the equations are
\begin{widetext}
\begin{subequations}
\begin{align}
\frac{\dif \bar{m}_1^{\rm out}}{\dif x}
=&
-\left(  x\frac{\dif \mathcal{\bar{R}}_0^{\rm out}}{\dif x}+4\mathcal{\bar{R}}_0^{\rm out}\right)  
\frac{\dif \bar{m}_0^{\rm out}}{\dif x} 
+\frac{x^2}{12}( 48\pi\bar{P}_1^{\rm out}
+3( \mathcal{\bar{R}}_0^{\rm out})^2+2\mathcal{\bar{R}}_1^{\rm out} ) 
+\frac{x^2}{6}
( 48\pi\bar{P}_0^{\rm out}+32\pi\bar{\rho}+2\mathcal{\bar{R}}_0^{\rm out}) 
\mathcal{\bar{R}}_0^{\rm out} 
\notag \\
&
+\frac{1}{6}
( 16\pi\bar{\rho}x^3+x^3\mathcal{\bar{R}}_0^{\rm out}-6\bar{m}_0^{\rm out} )
\mathcal{\bar{R}}_0^{\rm out \prime}, \label{Out_m1_dif}
\\
4x ( x-2\bar{m}_0^{\rm out} )\frac{\dif \bar{P}_1^{\rm out}}{\dif x}
=&
-[
4x( x-2\bar{m}_0^{\rm out} )
( 2\mathcal{\bar{R}}_0^{\rm out}-x\mathcal{\bar{R}}_0^{\rm out\,\prime} )
-8x\bar{m}_1^{\rm out}
]
\frac{\dif \bar{P}_0^{\rm out}}{\dif x}
-( \bar{P}_0^{\rm out}+\bar{\rho} )  
[ 16\pi x^3\bar{P}_1^{\rm out}-x^3( \mathcal{\bar{R}}_0^{\rm out} )^2
\notag \\
&
-8x (x-2\bar{m}_0^{\rm out})\mathcal{\bar{R}}_0^{\rm out\,\prime}+8\bar{m}_0^{\rm out}\mathcal{\bar{R}}_0^{\rm out}+4\bar{m}_1^{\rm out} ]
-\bar{P}_1^{\rm out}
( 16\pi x^3\bar{P}_0^{\rm out}+4\bar{m}_0^{\rm out}), \label{Out_P1_dif}
\\
x( x-2\bar{m}_0^{\rm out} ) 
\mathcal{\bar{R}}_0^{\rm out\,\prime\prime}
=&  \frac{x^2}{6}
( 24\pi\bar{P}_1^{\rm out} +\mathcal{\bar{R}}_1^{\rm out} ) 
+\frac{x^2}{3}
( 24\pi\bar{P}_0^{\rm out}-8\pi\bar{\rho}+\mathcal{\bar{R}}_0^{\rm out} ) 
\mathcal{\bar{R}}_0^{\rm out}
+2(\bar{m}_0^{\rm out}-x) 
\mathcal{\bar{R}}_0^{\rm out\,\prime}
+\frac{x^3}{6}
( 16\pi\rho+\mathcal{\bar{R}}_0^{\rm out} )  
\mathcal{\bar{R}}_0^{\rm out\,\prime}. \label{Out_R1_dif}
\end{align}
\end{subequations}
In Eq.\ \eqref{Out_m1_dif}, the $\mathcal{\bar{R}}_1^{\rm out}$ and $\bar{P}_1^{\rm out}$
terms can be eliminated by using Eq.\ \eqref{Out_R1_dif}. After that, the integration of the
equation gives the $O(\epsilon)$ outer solution of the mass
in terms of $\bar{P}_0^{\rm out}$ as
\begin{align}
\bar{m}_1^{\rm out}\left(x\right)=&
96\pi^2 x^3
\left(\bar{\rho}+\bar{P}_0^{\rm out}\right) 
\left(\frac{\bar{\rho}}{3}+\bar{P}_0^{\rm out}\right) 
-16\pi^2\bar{\rho}x^3
\left(4\bar{P}_0^{\rm out}+\bar{\rho}\right)
+\int \left(144\pi^2 x^2\bar{P}_0^{{\rm out} 2} 
+288\pi^2\bar{\rho}x^2\bar{P}_0^{\rm out}\right)
\,\dif x. \label{sol_out_m1}
\end{align}
Unfortunately, Eq.\ \eqref{Out_P1_dif} cannot be solved analytically. 
Still, the $O(\epsilon)$ outer solution of the Ricci 
scalar by using Eq.\ \eqref{Out_R1_dif} can be written in terms of the other outer solutions
as
\begin{align}
\mathcal{\bar{R}}_1^{\mathrm{out}}\left( x \right)=&
-144\pi\left(1-\frac{8}{3}\pi \bar{\rho}x^2\right)
\frac{\dif^2 \bar{P}_0^{\rm out}}{\dif x^2}
-\frac{48\pi}{x}
\left[ 
4\pi x^2\left(3\bar{P}_0^{\mathrm{out}}-5\bar{\rho}\right)+6 
\right]
\frac{\dif \bar{P}_0^{\mathrm{out}}}{\dif x}
-24\pi\bar{P}_1^{\rm out}.
\label{sol_out_R1}
\end{align}
\end{widetext}

\subsection{Inner solutions for uniform density}

Since the trace equation will be a second order differential equation
in the inner region despite it being an algebraic equation in the outer region,
it is a fair assumption that the boundary layer  occurs near the surface of the star
where the Ricci scalar changes its behavior to satisfy the boundary condition.
Therefore, we define the inner variable
(coordinate stretching parameter) as $\xi \equiv \left( 1-x\right)
/\epsilon^{\nu}$ where $\nu$ is to be determined by careful balancing of the terms.
Accordingly, writing Eq.\ \eqref{dif_cur} in terms of the inner variable, the $\mathcal{R}^{\prime\prime}$ term 
and one of the terms on the right-hand side of the equation become $O(1)$ 
while the rest of the terms are higher order. To obtain that we are forced to choose $\nu = 1/2$.

Hence, the inner solutions, valid for $0\ll x<1$, are introduced as
\begin{subequations}
\begin{flalign}
\bar{\mathcal{R}}^{\text{in}}(\xi )=& \bar{\mathcal{R}}_{0}^{\text{in}}\left( \xi \right)
+\epsilon^{1/2}\bar{\mathcal{R}}_{1}^{\text{in}}\left( \xi \right) 
+\epsilon \bar{\mathcal{R}}_{2}^{\text{in}}\left( \xi \right)
+O(\epsilon^{3/2}),& \\
\bar{m}^{\text{in}}(\xi )=& \bar{m}_{0}^{\text{in}}\left( \xi \right)
+\epsilon^{1/2}\bar{m}_{1}^{\text{in}}\left( \xi \right) 
+\epsilon\bar{m}_{2}^{\text{in}}\left(\xi\right) 
+O(\epsilon^{3/2}),& \\
\bar{P}^{\text{in}}(\xi )=& \bar{P}_{0}^{\text{in}}\left( \xi \right)
+\epsilon^{1/2}\bar{P}_{1}^{\text{in}}\left( \xi \right) 
+\epsilon\bar{P}_{2}^{\text{in}}\left( \xi \right) 
+O(\epsilon^{3/2}).&
\end{flalign}
\end{subequations}
They should satisfy the boundary conditions at $x=1$,
\begin{align}
&\bar{P}_{0}^{\mathrm{in}}(\xi=0)=\bar{P}_{1}^{\mathrm{in}}(\xi=0)=
\bar{\mathcal{R}}_{1}^{\mathrm{in}}(\xi=0)=0,
\notag \\
&\bar{\mathcal{R}}_{0}^{\mathrm{in}}(\xi=0)=8\pi\bar{\rho}.
\label{In_bcs}
\end{align}
\begin{widetext}
Using the coordinate stretching parameter we rewrite the dimensionless TOV Eq.\ \eqref{dif_m} as
\begin{align}
\frac{2}{\epsilon ^{1/2}}[ 
1+2\epsilon\bar{\mathcal{R}}^{\mathrm{in}}-\epsilon^{1/2}
( 1-\epsilon^{1/2}\xi)\mathcal{\bar{R}}^{\mathrm{in}\,\prime}] \frac{\dif\bar{m}^{\mathrm{in}}}{\dif\xi }=&
-\frac{\left( 1-\epsilon ^{1/2}\xi \right) ^{2}}{6}
\left( 48\pi \bar{P}^{\mathrm{in}}+\left( 2+3\epsilon \bar{\mathcal{R}}^{\rm in}\right) \bar{\mathcal{R}}^{\mathrm{in}}+32\pi 
\bar{\rho}\right)  
\notag \\
& +\epsilon^{1/2}\frac{ \mathcal{\bar{R}}^{\mathrm{in}\,\prime}}
{3( 1+2\epsilon \bar{\mathcal{R}}^{\mathrm{in}}) }[ 
( 1-\epsilon^{1/2}\xi)^{3}
( \bar{\mathcal{R}}^{\mathrm{in}}+3\epsilon ( \bar{\mathcal{R}}^{
\mathrm{in}})^{2}+16\pi \bar{\rho}) 
-6\bar{m}^{\mathrm{in}}( 1+2\epsilon \bar{\mathcal{R}}^{\mathrm{in}})]  \notag
\\
& -\epsilon \left( 1-\epsilon^{1/2}\xi \right) 
\left( 1-\epsilon^{1/2}\xi-2\bar{m}^{\mathrm{in}}\right) 
\frac{4}{\left( 1+2\epsilon \bar{\mathcal{R}}^{\mathrm{in}}\right)}
\left( \bar{\mathcal{R}}^{\mathrm{in}\,\prime}\right)^{ 2},
\label{In_m_dif}
\end{align}
TOV Eq.\ \eqref{dif_P} as
\begin{align}
&\left[ 
4\left( 1-\epsilon ^{1/2}\xi \right) \left( 1-\epsilon ^{1/2}\xi -2\bar{m}^{\mathrm{in}}\right) \left( 1+2\epsilon \bar{\mathcal{R}}^{\mathrm{in}}-\left(
\epsilon ^{1/2} -\epsilon \xi \right) \left( \bar{\mathcal{R}}^{\mathrm{in}}\right)
^{\prime }\right) \right]\frac{\dif\bar{P}^{\mathrm{in}}}{\dif\xi }\notag \\
&\p{a}=  
\epsilon ^{1/2} ( \bar{\rho}+\bar{P}^{\mathrm{in}}) 
[ 
16\pi ( 1-\epsilon^{1/2}\xi)^{3}\bar{P}^{\mathrm{in}}
+4\bar{m}^{\mathrm{in}}+8\epsilon\bar{m}^{\mathrm{in}}\bar{\mathcal{R}}^{\mathrm{in}} 
-\epsilon ( 1-\epsilon^{1/2}\xi )^{3} (\bar{\mathcal{R}}^{\mathrm{in}} )^{2} 
+8\epsilon^{1/2} ( 1-\epsilon^{1/2}\xi ) 
( 1-\epsilon ^{1/2}\xi -2\bar{m}^{\mathrm{in}}) 
\bar{\mathcal{R}}^{\mathrm{in}\,\prime } 
], \label{In_P_dif}
\end{align}%
and finally the trace Eq.\ \eqref{dif_cur} as
\begin{align}
&\left( 1-\epsilon ^{1/2}\xi \right)\left( 1-\epsilon ^{1/2}\xi -2\bar{m}^{\mathrm{in}} \right) \left( 1+2\epsilon \bar{\mathcal{R}}^{\mathrm{in} }\right) \bar{\mathcal{R}}^{\rm in\,\prime\prime}  \notag \\
&\p{a}=\frac{1}{6}\left( 1-\epsilon ^{1/2}\xi \right) ^{2}\left( 1+2\epsilon 
\bar{\mathcal{R}}^{ \mathrm{in}}\right) \left(-8\pi \bar{\rho}+24\pi 
\bar{P}^{\mathrm{in}}+\bar{\mathcal{R}}^{\mathrm{in}}\right)  \notag \\
&\p{aa} -\frac{\epsilon ^{1/2}}{6}\left[ \left( 1+2\epsilon \bar{\mathcal{R}}^{\mathrm{in}%
}\right) 12\bar{m}^{\mathrm{in}}-12\left( 1+2\epsilon \bar{\mathcal{R}}^{\mathrm{in}%
}\right) \left( 1-\epsilon ^{1/2}\xi \right) +\epsilon \left( 1-\epsilon^{1/2}\xi \right)^{3}\left( \bar{\mathcal{R}}^{\mathrm{in}}\right)^{2}
\right]  \bar{\mathcal{R}}^{\mathrm{in}\,\prime }  \notag \\
&\p{aa} -\frac{\epsilon ^{1/2}}{6} \left[ \left( 1-\epsilon ^{1/2}\xi \right)^{3}%
\bar{\mathcal{R}}^{\mathrm{in} }+16\pi \left( 1-\epsilon^{1/2}\xi \right)^{3}\bar{\rho} 
\right]\bar{\mathcal{R}}^{\mathrm{in}\,\prime} 
+2\epsilon \left( 1-\epsilon ^{1/2}\xi \right) \left( 1-\epsilon ^{1/2}\xi-2\bar{m}^{\mathrm{in}} \right) 
\left(\bar{\mathcal{R}}^{\mathrm{in}\,\prime}\right)^{2}.  
\label{In_R_dif}
\end{align}%
$O(1) $ terms in these equations are
\begin{subequations}
\begin{align}
\frac{\dif\bar{m}_{0}^{\mathrm{in}}}{\dif\xi }=&0, \\
\left( 1-2\bar{m}_{0}^{\mathrm{in}}\right) \frac{\dif\bar{P}_{0}^{\mathrm{in}}}{%
\dif\xi }=&0, \\
6\left( 1-2\bar{m}_{0}^{\mathrm{in}}\right) \frac{ {\rm d^2}\bar{\mathcal{R}}_{0}^{\rm in}}{ {\rm d}\xi^2  }=&-8\pi \bar{\rho}+24\pi \bar{P}
_{0}^{\mathrm{in}}+\bar{\mathcal{R}}_{0}^{\mathrm{in}}.  \label{in_R0_dif}
\end{align}
\end{subequations}
\end{widetext}
The nontrivial solutions of these equations are
\begin{align}
&\bar{m}_{0}^{\rm in}\left( \xi \right)=A_0, \qquad 
\bar{P}_{0}^{\rm in}\left( \xi \right)=0, 
\notag \\
&\bar{\mathcal{R}}_{0}^{\mathrm{in}}\left( \xi \right) =C_{0}\exp \left( \frac{\xi }{%
\sqrt{6\left( 1-2A_{0}\right) }}\right) 
\notag \\
&+D_{0}\exp \left( -\frac{\xi }{\sqrt{%
6\left( 1-2A_{0}\right) }}\right) +8\pi \bar{\rho},
\end{align}%
where $C_{0}+D_{0}=0$, 
with the boundary conditions given in Eq.\ \eqref{In_bcs}.
$C_0$ should be zero to prevent infinities at the matching procedure.
So, $D_0$ is also zero.

Then, $O(\epsilon ^{1/2})$ terms in the equations of the inner solution [Eqs.\ \eqref{In_m_dif}, \eqref{In_P_dif}, and \eqref{In_R_dif}] are
\begin{subequations}
\begin{align}
\frac{\dif\bar{m}_{1}^{\mathrm{in}}}{\dif\xi }=& 
-\frac{1}{6}\left( \bar{\mathcal{R}}_{0}^{\mathrm{in}}+16\pi \bar{\rho}\right)  ,
\label{In_m1_dif} \\
\left( 1-2A_{0}\right) \frac{\dif\bar{P}_{1}^{\mathrm{in}}}{\dif\xi }=& 
\bar{\rho}A_{0},
\label{In_P1_dif} \\
\left( 1-2A_{0}\right) \frac{ {\rm d^2}\bar{\mathcal{R}}_{1}^{\rm in}}{ {\rm d}\xi^2 }=
& \frac{1}{6}\left( 24\pi \bar{P}_{1}^{%
\mathrm{in}}+\bar{\mathcal{R}}_{1}^{\mathrm{in}}\right),
\label{In_R1_dif}
\end{align}
\end{subequations}
and the solution of these equations 
with the boundary conditions given in Eq.\ \eqref{In_bcs}
are
\begin{subequations}
\begin{align}
\bar{m}_{1}^{\mathrm{in}}\left( \xi \right) =
& -4\pi \bar{\rho}\xi +A_{1}, 
\\
\bar{P}_{1}^{\mathrm{in}}\left( \xi \right) =& 
\frac{A_{0}\bar{\rho}}{\left( 1-2A_{0}\right) }\xi,
\\
\bar{\mathcal{R}}_{1}^{\mathrm{in}}\left( \xi \right) =
& C_{1}\exp \left( \frac{\xi }{\sqrt{6-12A_{0}}}\right) \notag \\
&+D_{1}\exp \left( -\frac{\xi }{\sqrt{6-12A_{0}}}\right) 
-\frac{24\pi \bar{\rho}A_{0}}{1-2A_{0}}\xi,
\end{align}
\end{subequations}
where $C_1+D_1=0$. Again, to prevent infinities in a matching procedure
$C_1$ should be zero. So, $D_1$ is also zero.

After applying these solutions,
$O(\epsilon)$ terms in the equations of the inner solution [Eqs.\ \eqref{In_m_dif}, \eqref{In_P_dif} and \eqref{In_R_dif}] are
obtained as
\begin{subequations}
\begin{gather}
\frac{\dif \bar{m}_2^{\rm in}}{\dif \xi}= 8\pi\bar{\rho}\xi, 
\\
\frac{ {\rm d}\bar{P}_2^{\rm in} }{ {\rm d}\xi }
= -\frac{\bar{\rho}}{\left( 1-2A_0 \right)^2}
\left[ 4\pi\bar{\rho}\left(1-A_0\right)\xi-A_0\left(2-A_0\right)\xi -A_1\right],
\\
(1-2A_0) \frac{\dif^2 \mathcal{\bar{R}}_2^{\rm in}}{\dif \xi^2}
=
\frac{1}{6}(24\pi\mathcal{\bar{P}}_2^{\rm in} +\mathcal{\bar{R}}_2^{\rm in})
+\frac{48\pi\bar{\rho}}{1-2A_0}(2\pi\bar{\rho}+A_0-1).
\end{gather}
\end{subequations}
\begin{widetext}
Accordingly, $O(\epsilon)$ inner solutions are
\begin{subequations}
\begin{align}
\bar{m}_2^{\rm in}\left(\xi\right) = &
4\pi\bar{\rho}\xi^2+A_2,
\\
\bar{P}_2^{\rm in}\left(\xi\right) = &
-\frac{\bar{\rho}}{\left(1-2A_0\right)^2}
\left(
2\pi\xi^2\bar{\rho}\left(1-A_0\right)
-\frac{1}{2}A_0\xi^2\left(2-A_0\right)
-A_1\xi
\right),
\\
\mathcal{\bar{R}}_2^{\rm in}\left(\xi\right) = &
C_2\exp\left(
\frac {\xi}{\sqrt{6\left(1-2A_0\right)}}
\right)
+D_2\exp\left(
-\frac {\xi}{\sqrt{6\left(1-2A_0\right)}}
\right)
+\frac{12\pi\bar{\rho}}{\left( 1-2A_0 \right)^2}
\xi^2\left[ 
A_0^2-A_0\left(4\pi\bar{\rho}+2\right)
+4\pi\bar{\rho}
\right]
\notag \\
&-\frac{24A_1\pi\bar{\rho}}
{\left(1-2A_0\right)^2}\xi 
+\frac{288\pi\bar{\rho}}{\left(1-2A_0\right)^2}
\left[
A_0^3+\left(8\pi\bar{\rho}-\frac{1}{2}\right)A_0^2
-8\pi\bar{\rho}A_0+2\pi\bar{\rho}
\right].
\end{align}
\end{subequations}
Again, $C_2$ should be zero to prevent infinities in the 
matching procedure. Then, the boundary condition,\newline $\mathcal{R}_2^{\rm in}\left(\xi=0\right)=0$, implies 
that
\begin{equation}
D_2=
-\frac{288\pi\bar{\rho}}{\left(1-2A_0\right)^2}
\left[
A_0^3+\left(8\pi\bar{\rho}-\frac{1}{2}\right)A_0^2
-8\pi\bar{\rho}A_0+2\pi\bar{\rho}
\right].
\end{equation}
\end{widetext}
\subsection{Composite solutions for uniform density}

For matching the solutions we employed Van Dyke's method.
As shown in Sec.\ \ref{sec:match}, the solutions
can match and we obtain
\begin{gather}
P_c = \bar{\rho}\frac{1-\sqrt{1-2A_0}}{3\sqrt{1-2A_0}-1}, \quad 
A_0=\frac{4}{3}\pi\bar{\rho}, \notag \\
\quad \bar{P}_2^{\rm out}\left(x=1\right)=A_1=0, \quad A_2=\bar{m}_1^{\rm out}\left(x=1\right).
\end{gather}

\begin{figure*}
\centering
\includegraphics[width=0.48\textwidth]{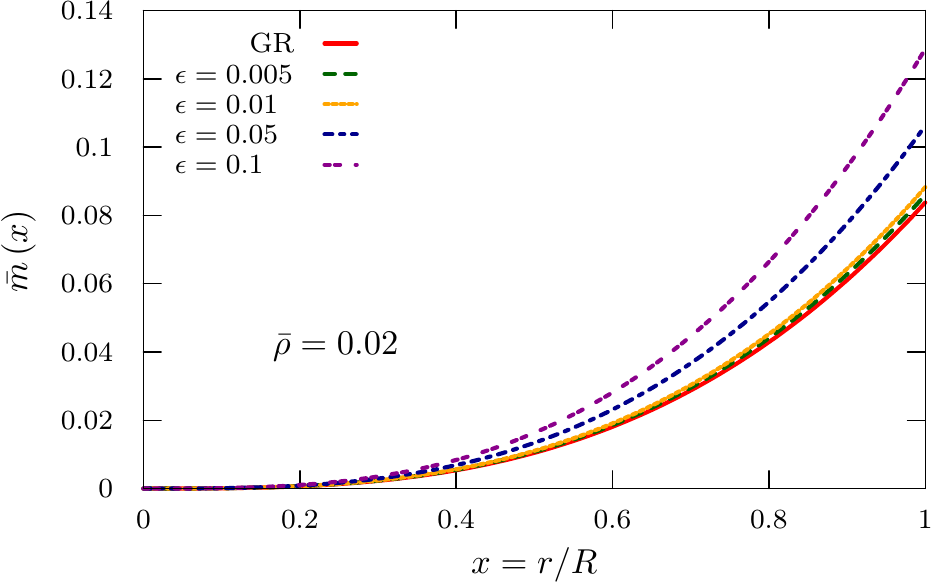}
\includegraphics[width=0.48\textwidth]{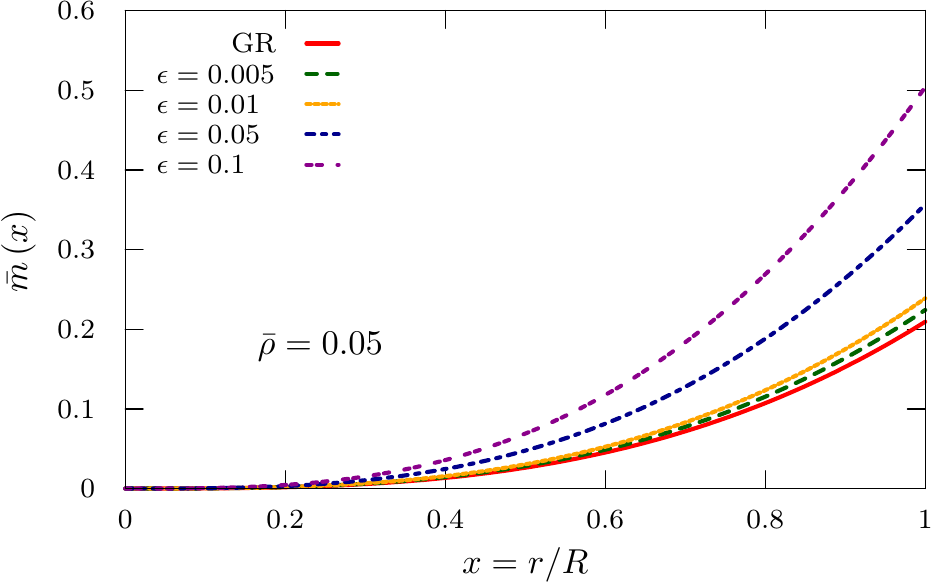}
\caption{Dimensionless mass distribution inside of the star with dimensionless density of $0.02$ and $0.05$ for various values of $\epsilon$.}
\label{fig:mr}
\end{figure*}

After matching the solutions we can construct the composite solutions by
subtracting the overlapping parts from the sum of the solutions. Accordingly,
the dimensionless composite solutions are
\begin{subequations}
\begin{align}
\bar{m}^{\rm comp}\left(x\right)=&\bar{m}_0^{\rm out}+\epsilon\bar{m}_2^{\rm out},
\label{uni_mass} \\
\bar{P}^{\rm comp}\left(x\right)=&\bar{P}_0^{\rm out}+\epsilon\bar{P}_2^{\rm out},
\label{uni_Pressure} \\
\bar{R}^{\rm comp}\left(x\right)=&\bar{R}_0^{\rm out}+\epsilon\bar{R}_2^{\rm out}
+\epsilon\frac{108A_0^2}{1-2A_0}\left(3-7A_0\right)
\notag \\
&\times\left[1-
\exp\left(-\frac {1-x}{\sqrt{6\epsilon\left(1-2A_0\right)}}
\right)
\right]. \label{uni_Ricci}
\end{align} 
\end{subequations}
The dimensionless Ricci scalar has a contribution from
the inner solution. Yet,
the dimensionless mass and the pressure are the same as the
regular perturbation approach. 
So, they do not change their behaviors near the surface of the star
as shown in Fig.\ \ref{fig:mr} for the dimensionless mass. This result might be caused 
by our assumption of uniform density or the boundary condition on the Ricci scalar at the surface of the star.

\section{Mass-Radius Relation}
\label{sec:mr}

\begin{figure}[h!]
\centering
\includegraphics[width=0.48\textwidth]{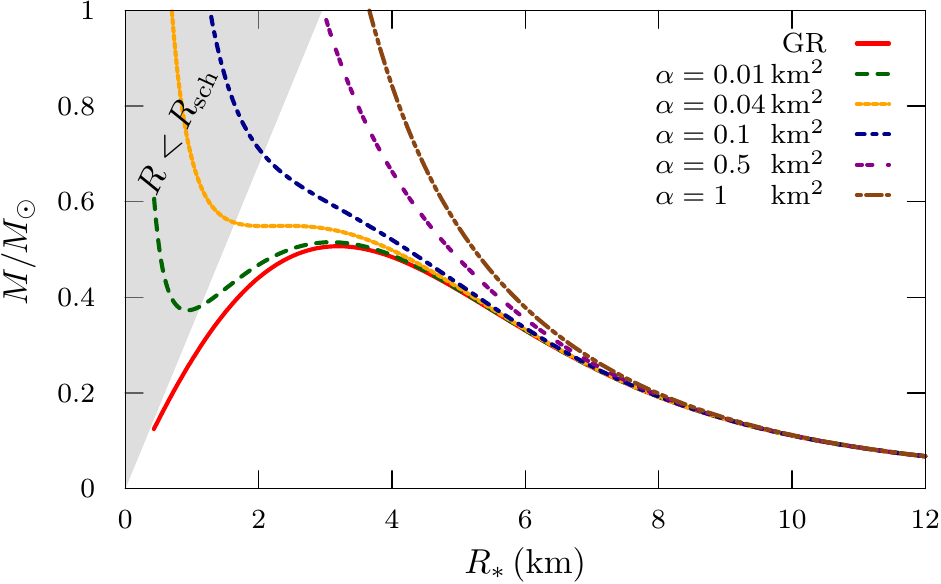}
\caption{The stellar mass and the radius of the star (M-R) relations for different values of $\protect\alpha$. Here $M$, $%
M_\odot$, and $R_\ast$ are the stellar mass, the Sun mass, and radius of the
star, respectively. The grey region is where the radius of the star is smaller than the Schwarzschild radius.}
\label{fig:MR1}
\end{figure}

\begin{figure*}
\centering
\includegraphics[width=0.48\textwidth]{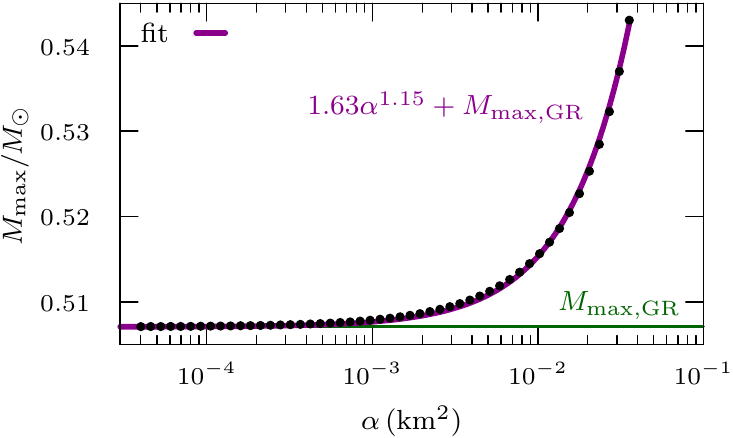}
\includegraphics[width=0.48\textwidth]{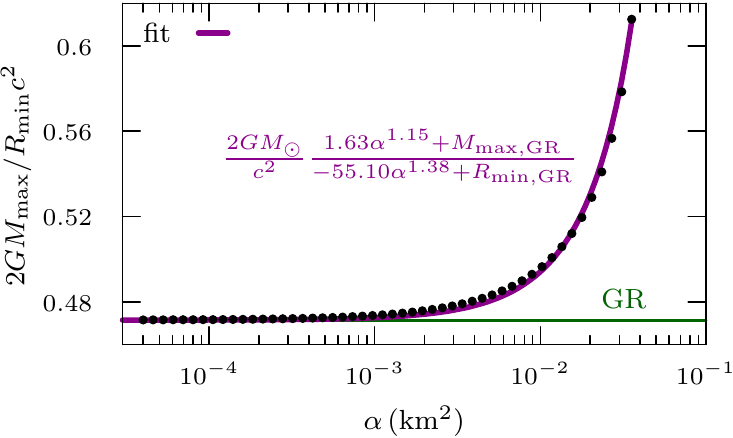}
\caption{Left: Maximum mass of the star vs $\alpha$. Right: Compactness which corresponds to the maximum mass of the star vs $\alpha$.
Here $M_\odot$ is the solar mass.
\label{fig:MR_fit}
}
\end{figure*}

All solutions and parameters can
be written in dimensional form by referring to Eq.\ \eqref{nondim}. Then, the radius of the star can be
found in terms of the relativity parameter, $\mu
=\left(9\pi/4\right)^{1/3}\left(\rho/\rho_{L}\right)^{1/3}$, as
\begin{equation}
R(\mu) = R_{L}\left( \frac{2}{5\mu}\right)^{1/2} \left( \frac{9\pi }{4}
g\left( \mu\right)\right)^{1/3}, \label{Radius_s}
\end{equation}
where 
\begin{equation}
g\left( \mu\right) =\frac{\left( 1+2\mu^{2}/5\right) ^{3/2}}{\left(
1+3\mu^{2}/5\right) ^{3}}
\end{equation}
and the Landau parameters are 
\begin{align}
M_L \equiv& \frac{1}{m_N^2}\left(\frac{\hbar c}{G}\right)^{3/2}, \quad
R_{L} \equiv \frac{GM_{L}}{c^{2}} \approx 2.71 \, \mathrm{km}, 
\notag \\
\rho _{L} \equiv& \frac{3M_{L}}{4\pi R_{L}^{3}}\approx 4.29\times 10^{16}\,
\mathrm{g\,cm}^{-3}
\end{align}
\cite{ghosh07}. By using these, the mass-radius (M-R) relation can be
obtained as shown in Fig.\ \ref{fig:MR1}. We see that
the mass of the star
increases with $\alpha$ for a star with a fixed radius.
The solutions with $dM/d\rho<0$ are unstable, a well known result from GR, which we assume to prevail in $f({\cal R})$ gravity. 
The maximum mass is achieved when $dM/d\rho=0$.
The model thus predicts greater maximum mass, $M_{\max}$, depending on the value of $\alpha$. As shown in the figure, for the nonzero values of $\alpha$ we find solutions for which $dM/d\rho>0$ beyond densities exceeding the one yielding $M_{\max}$. Yet,
these solutions cannot be stable as they occur when the object is totally contained within the Schwarzschild radius. When $\alpha$ is greater than $0.04\,{\rm km^2}$,
the unstable branch vanishes and we cannot determine the maximum mass as in GR. Hence, unlike GR, the stable solutions go to Schwarzschild solution continuously. For lower values of $\alpha$ than $0.04\,{\rm km^2}$, the maximum mass of the stable star and the corresponding compactness are shown in Fig.\ \ref{fig:MR_fit}. Accordingly, the maximum mass and the compactness can be represented, when $\alpha<0.04\,{\rm km^2}$, with
\begin{gather}
\frac{M_{\rm max}}{M_\odot}=1.63\alpha^{1.15}+M_{\rm max,GR},
\notag \\
\frac{2GM_{\rm max}}{R_{min} c^2}=\frac{2GM_\odot}{c^2}\frac{1.63\alpha^{1.15}+M_{\rm max,GR}}{-55.19\alpha^{1.38}+R_{\rm min,GR}},
\end{gather}
where $M_\odot$ is the solar mass, $M_{\rm max,GR}$ and $R_{\rm min,GR}$ are the mass and the radius of the highest mass star in GR, and their values are, respectively, $0.5M_\odot$ and $3.19\, {\rm km}$.

\section{Conclusion}
\label{sec:conc}

We studied the structure and mass-radius relation for uniform density relativistic stars in the $f(\mathcal{R}) = \mathcal{R}+\alpha \mathcal{R}^{2}$ gravity model. We used the method of matched asymptotic expansions to handle the singular perturbation problem posed by the higher order derivatives in the field equations arising from the higher order curvature term. This method allows us to obtain solutions, parametrized by $\alpha$, which smoothly match with the solutions obtained in the general relativity ($\alpha=0$). This establishes, once again by a different method than the previously considered perturbative approach \citep{Psaltis+10,Eksi+11}, the existence of relativistic stars in this model of gravity.
The solutions of the mass and the pressure obtained in this paper are 
the same as the regular perturbative approach since their composite solutions contain only outer solutions. So, the MAE method does not provide a different mass-radius relation
in this case. Yet, this outcome might change with a different choice of the boundary condition for the Ricci scalar at the surface of the star.

We find that $M_{\max}$ increases almost linearly with $\alpha$ while $R$ decreases with $\alpha^{-1.38}$.
This implies that general relativity, as a special case of this model of gravity, holds the least compact stellar configurations.
References \cite{Eksi+11,Yaz+14a} find that the maximum mass, $M_{\max}$, has a minimum value at a certain value of $\alpha \sim 1.5\times 10^{10}\,{\rm cm^2}$ and  $\alpha \sim 1\times 10^{10}\,{\rm cm^2}$, respectively, for the polytropic equation state with polytropic index of $9/5$.
The difference between the result in this paper and the previous findings likely stems from the uniform density assumption employed in this paper
which leads to the vanishing of terms involving $d\rho/dr$.

As a future work, the same calculations should be repeated for realistic equation of states
with the MAE method by permitting the vacuum solutions other than Schwarzschild's solution, and the results should be
compared with the previous works.

\appendix

\section{SOLUTION OF $\bar{P}_{0}^{\mathrm{out}}$ \label{app_uni_P0}}

With Eq.\ \eqref{Uni_m0}, Eq.\ \eqref{out_P_0_diff} can be written as 
\begin{equation}
\frac{\dif\bar{P}_{0}^{\mathrm{out}}}
{\left( \bar{\rho}+\bar{P}_{0}^{\mathrm{out}}\right) 
\left( \bar{P}_{0}^{\mathrm{out}}+\bar{\rho}/3\right) }=
-4\pi \frac{x\dif x}{1-\frac{8}{3}\pi x^{2}\bar{\rho} }.
\end{equation}%
By integrating both sides from $\bar{P}_c$ to $\bar{P}_{0}^{\mathrm{out}}$ and from $0$ to $x$
\begin{equation}
\ln \left( \frac{\left( 3\bar{P}_{0}^{\mathrm{out}}+\bar{\rho}\right) \left( \bar{P}_{c}+\bar{\rho}\right)}
{\left( \bar{P}_{0}^{\mathrm{out}}+\bar{\rho}\right) 
\left( 3\bar{P}_{c}+\bar{\rho}\right) }\right)^{3/2\bar{\rho}}
=\ln \left( \frac{8\pi \bar{\rho}x^2-3}{-3}\right)^{3/4\bar{\rho}}
\end{equation}%
is obtained. Solving this equation we find
\begin{equation}
\bar{P}_{0}^{\mathrm{out}}=\bar{\rho}\left[ 
\frac{2\left(P_{c}+\bar{\rho}\right)}
{3\left( P_{c}+\bar{\rho}\right) 
-\left(3P_{c}+\bar{\rho}\right)
\sqrt{1-\frac{8}{3}\pi \bar{\rho}x^2} 
}-1\right],
\end{equation}%
where $\bar{P}_{c}$ is a constant corresponding to $\bar{P}_{0}^{\mathrm{out}}\left(
0\right) $.

\section{MATCHING THE SOLUTIONS \label{sec:match}}
According to Van Dyke's method, first the outer solutions are written 
in terms of the inner variable, and they are expanded  up to $O(\epsilon)$
for small $\xi$. Similarly, the inner solutions are written in terms of
$x$, and they are expanded up to $O(\epsilon)$
for small $\epsilon$. Then, the matching conditions are 
obtained by equaling both of them.

Accordingly, the matching condition for the mass solutions is
\begin{align}
&\frac{4}{3}\pi \bar{\rho}\left(1-3x+3x^2\right)
+\epsilon \bar{m}_1^{\rm out}\left(1\right)
\notag \\
&\p{a}=A_0-4\pi\bar{\rho}\left(1-x\right)+\sqrt{\epsilon}A_1+4\pi\bar{\rho}\left(1-x\right)^2
+\epsilon A_2.
\end{align}
Obviously, $A_1$ should be zero and
\begin{align}
\frac{4}{3}\pi \bar{\rho}=A_0, \quad
\bar{m}_1^{\rm out}\left(1\right)=A_2.
\end{align}

Similarly, the matching condition for solutions of the pressure is
\begin{widetext}
\begin{align}
\bar{\rho}&\left[\frac{2K}{3K-L\sqrt{1-2\beta}}-1\right]
+\frac{4KL\bar{\rho}\beta\left(1-x\right)}
{\left(3K-L\sqrt{1-2\beta}\right)^2\sqrt{1-2\beta}}
+\frac{2\bar{\rho}}{3K-L\sqrt{1-2\beta}} 
\left[
\frac{LK\sqrt{1-2\beta}}{3K-L\sqrt{1-2\beta}}
\left( 
-\frac{\beta}{1-2\beta}
-\frac{2\beta^2}{\left(1-2\beta\right)^2}
\right) 
\right.
\notag \\
&\left.
+\frac{4KL^2\beta^2}{ 
\left(3K-L\sqrt{1-2\beta}\right)^{2} 
\left(1-2\beta\right) 
} \right]\left(1-x\right)^2
+\epsilon \bar{P}_2^{\rm out}\left(1\right)= \frac{A_{0}\bar{\rho}}{2\left(1-2A_0\right)}
\left(1-x^2\right), \label{match_Press}
\end{align}
\end{widetext}
where
\begin{equation}
\beta=\frac{4}{3}\pi\bar{\rho}, \quad
K=P_c+\bar{\rho}, \quad
L=3P_c+\bar{\rho}.
\end{equation}
Obviously, $\bar{P}_2^{\rm out}\left(1\right)$ should be zero. 
The right-hand side (RHS) and left-hand side (LHS) of Eq.\  \eqref{match_Press} can be matched if the following conditions 
are satisfied:
\begin{enumerate}[(i)]
\item The third term equals the opposite sign and half of the second term in LHS of the equation
since there is not any term proportional to $x$ in the RHS of the equation.
\item The summation of factors of the first, second, and third terms 
equals the opposite sign of the factor of the third term in the LHS of the equation
since the factor of $x^2$ is the opposite sign of the constant term in the RHS of the equation. 
\item Combining the first condition with the second condition gives that
the first term in the LHS of the equation should be zero.
\end{enumerate}
All conditions are satisfied if
\begin{equation}
P_c = \bar{\rho}\frac{1-\sqrt{1-2A_0}}{3\sqrt{1-2A_0}-1}.
\end{equation}

Finally, by using the above results, the matching condition for the inner solution of the Ricci scalar can be written as
\begin{align}
&6A_0
-\frac{18A_{0}^2}{1-2A_{0}}
\left(1-x\right)
+\frac{9A_0}{\left( 1-2A_0 \right)^2}
\left[ 
-2A_0^2+A_0
\right] 
\left(1-x\right)^2
\notag \\
&\p{a}+\epsilon\frac{108A_0^2}{1-2A_0}
\left[3-7A_0\right], \label{match_Rin}
\end{align}
and for the outer solution of the Ricci scalar can be written as
\begin{align}
&6A_0-\frac{9A_0^2}{1-2A_0}
\left(1-x^2\right)
+24\pi\epsilon\left\lbrace
-6\left(1-2A_0\right)
\left.\frac{\dif^2 \bar{P}_0^{\rm out}}{\dif x^2}\right|_{x=1}
\right.
\notag \\
&\p{a}
\left.
+
\left(30A_0-12\right)
\left.\frac{\dif \bar{P}_0^{\rm out}}{\dif x}\right|_{x=1}
\right\rbrace, \label{match_Rout}
\end{align}
and
\begin{align}
\left.\frac{\dif^2 \bar{P}_0^{\rm out}}{\dif x^2}\right|_{x=1}
=\left.\frac{\dif \bar{P}_0^{\rm out}}{\dif x}\right|_{x=1}
=-\frac{A_0}{1-2A_0}.
\end{align}
It can easily be shown that Eqs.\ \eqref{match_Rin} and
\eqref{match_Rout} are equal to each other.

\bibliography{ref}
\bibliographystyle{apsrev4-1}

\end{document}